\documentclass{article}
\usepackage{amsfonts}
\usepackage{amssymb}
\usepackage{amsmath}
\usepackage{graphicx}
\usepackage{float}
\usepackage[top=0.75in, bottom=0.75in, left=1.0in, right=1.0in]{geometry}

\begin{document}

\title{New solution to quantum master equation for diffusion\thanks{{\small %
Work supported by the National Natural Science Foundation of China under
grant 11175113 and 11264018.}}}
\author{Gao Fang$^{1}$, Fan Hong-yi$^{1}$, Shuang Feng$^{1,2}$, and Tang
Xu-bing$^{1,3\dag }$ \\
$^{1}$Institute of Intelligent Machines, Chinese Academy of Sciences,\\
Hefei 230031, China\\
$^{2}$Department of Automation, University of Science \& Technology\\
of China, Hefei 230027, China\\
$^{3}$School of Mathematics \& Physics Science and Engineering, \\
Anhui University of Technology, Ma'anshan 243032, China\\
$^{\dag }$ttxxbb@ahut.edu.cn $\thanks{%
Corresponding author.}$}
\maketitle

\begin{abstract}
Based on the Kraus-form solution to the master equation describing diffusion
we develop an integral-form solution by using the method of integration
within ordered product of operators, i.e., the evolution law of density
operator in diffusion channel can be considered as an integration
transformation from an input to its output density operator. It brings much
convenience for obtaining the time evolution law in the diffusion process
via this formalism.
\end{abstract}

PACS numbers: 42.50.Dv, 03.67. a, 42.50.Ex

\section{Introduction}

Due to the interaction between quantum system and environment, quantum
decoherence or quantum diffusion are inevitable \cite%
{Preskill_1998_lnp,Gardiner_2000_qn}. Quantum decoherence phenomena (QDP),
associated with quantum entanglement, teleportation, quantum noise, and
fluctuations, is an important issue received much attention by scientis.
Dynamics of this model can be described by the master equation or the
associated Langevin or Fokker-Planck equations \cite{Gardiner_2000_qn}. If
these systems involve negligible correlations amongst their components,
quantum memory (non-Markovian) effects can ignored. In other words,
Markovian approximation is valid. Considering a quantum harmonic system
interacted with a thermal bath, its master equation describing time
evolution of system's $\rho $ in the diffusion channel is given by \cite%
{Carmichael_1999_mefp,Carmichael_2008_ncf}%
\begin{equation}
\frac{d\rho }{dt}=-\kappa \left[ a^{\dagger }a\rho -a^{\dagger }\rho a-a\rho
a^{\dagger }+\rho aa^{\dagger }\right] ,  \label{1}
\end{equation}%
where $a^{\dagger }$ is the Bose creation operator satisfying the relation $%
\left[ a,a^{\dagger }\right] =1,$ and $\kappa $ is a diffusion constant. In
general, operator solutions to master equation are required to be in the
infinite sum representation (named Kraus formal solution) 
\begin{equation}
\rho \left( t\right) =\sum_{n=0}^{\infty }M_{n}\rho _{0}M_{n}^{\dagger },
\label{2}
\end{equation}%
where $\rho _{0}$ is the initial density operator, $M_{n}$ is the Kraus
operator \cite{Kraus_1983_lnp}, and satisfies the normalized condition, i.e, 
$\sum\limits_{n=0}^{\infty }M_{n}^{\dagger }M_{n}=1$. For different master
equations, there correspond to different forms of Kraus operator solutions.
However, the solutions in infinite sum representation to some master
equations have still not been found. In Ref. \cite{Fan_2008_mplb} by using
the entangled state representation and the method of integration within
ordered product (IWOP) of operators, the solution to Eq. (1) is derived 
\begin{eqnarray}
\rho \left( t\right)  &=&\sum_{m,n=0}^{\infty }\sqrt{\frac{1}{m!n!}\frac{%
\left( \kappa t\right) ^{m+n}}{\left( \kappa t+1\right) ^{m+n+1}}}a^{\dagger
m}\left( \frac{1}{1+\kappa t}\right) ^{a^{\dagger }a}a^{n}\rho
_{0}a^{\dagger n}\left( \frac{1}{1+\kappa t}\right) ^{a^{\dagger }a}a^{m}%
\sqrt{\frac{1}{m!n!}\frac{\left( \kappa t\right) ^{m+n}}{\left( \kappa
t+1\right) ^{m+n+1}}}  \notag \\
&=&\sum_{m,n=0}^{\infty }M_{m,n}\rho _{0}M_{m,n}^{\dagger },  \label{3}
\end{eqnarray}%
where%
\begin{equation}
M_{m,n}=\sqrt{\frac{1}{m!n!}\frac{\left( \kappa t\right) ^{m+n}}{\left(
\kappa t+1\right) ^{m+n+1}}}a^{\dagger m}\left( \frac{1}{1+\kappa t}\right)
^{a^{\dagger }a}a^{n},  \label{4}
\end{equation}%
satisfying $\sum\limits_{m,n=0}^{\infty }M_{m,n}^{\dagger }M_{m,n}=1$, which
is trace conservative. Even we have had the general form solution to the
diffusion equation, it still remains difficulty when $\rho _{0}$ is
complicated, because performing sum for various Bose operators on the right
hand side of Eq.(\ref{3}) is tough, since $a^{\dagger }$ is not commutative
with $a$. To avoid this difficulty, in this paper, we will employ the IWOP
method to develop the solution in Eq. (\ref{3}) into a new integration
transformation form. It brings much convenience for obtaining the time
evolution law in the diffusion process via this new formalism and can help
us to understand quantum diffusion more deeply.

\section{Integration-form solution to the dissipative master equation (form
1)}

In order to improve Eq.(\ref{3}), we introduce coherent state representation 
\cite{Klauder_1985_cs} $|\alpha \rangle =\exp \left[ -\frac{|\alpha |^{2}}{2}%
+\alpha a^{\dagger }\right] \left\vert 0\right\rangle ,$ satisfying the
eigenequation $a|\alpha \rangle =\alpha |\alpha \rangle ,$ whose
completeness relation is 
\begin{equation}
\int \frac{d^{2}\alpha }{\pi }|\alpha \rangle \left\langle \alpha
\right\vert =1.  \label{5}
\end{equation}%
Using the normally ordering form of vacuum projector $\left\vert
0\right\rangle \left\langle 0\right\vert =\colon e^{-a^{\dagger }a}\colon $
and the IWOP method, we can rewrite Eq.(\ref{5}) as \cite{Fan_2003_jopb}%
\begin{equation}
\int \frac{d^{2}\alpha }{\pi }|\alpha \rangle \left\langle \alpha
\right\vert =\int \frac{d^{2}\alpha }{\pi }\colon e^{-|\alpha
|^{2}+a^{\dagger }\alpha +a\alpha ^{\ast }-a^{\dagger }a}\colon =1.
\label{6}
\end{equation}%
Noting that any operator can be expanded according to coherent states
(so-called $P$-representation) \cite{Sudanshan_1963_prl}, i.e.,%
\begin{equation}
\rho _{0}=\int \frac{d^{2}\alpha }{\pi }P(\alpha ,0)|\alpha \rangle \langle
\alpha |,  \label{7}
\end{equation}%
using%
\begin{equation}
\left( \frac{1}{1+\kappa t}\right) ^{a^{\dagger }a}\left\vert \alpha
\right\rangle =e^{-a^{\dag }a\ln \left( 1+\kappa t\right) }\left\vert \alpha
\right\rangle =e^{-|\alpha |^{2}/2+\alpha a^{\dag }\frac{1}{1+\kappa t}%
}\left\vert 0\right\rangle ,  \label{8}
\end{equation}%
and substituting Eq.(\ref{7}) into Eq.(\ref{3}), we have%
\begin{eqnarray}
\rho \left( t\right)  &=&\sum_{m,n=0}^{\infty }\frac{1}{m!n!}\frac{\left(
\kappa t\right) ^{m+n}}{\left( \kappa t+1\right) ^{m+n+1}}a^{\dagger m} 
\notag \\
&&\times \left( \frac{1}{1+\kappa t}\right) ^{a^{\dagger }a}\int \frac{%
d^{2}\alpha }{\pi }P(\alpha ,0)\left\vert \alpha \right\vert ^{2n}|\alpha
\rangle \langle \alpha |\left( \frac{1}{1+\kappa t}\right) ^{a^{\dagger
}a}a^{m}  \label{9} \\
&=&\int \frac{d^{2}\alpha }{\pi }P(\alpha ,0)\sum_{m,n=0}^{\infty }\frac{%
\left\vert \alpha \right\vert ^{2n}}{m!n!}\frac{\left( \kappa t\right) ^{m+n}%
}{\left( \kappa t+1\right) ^{m+n+1}}a^{\dagger m}e^{-|\alpha |^{2}+\alpha
a^{\dag }\frac{1}{1+\kappa t}}\left\vert 0\right\rangle \langle 0|e^{\alpha
^{\ast }a\frac{1}{1+\kappa t}}a^{m}.  \notag
\end{eqnarray}%
Then using $\left\vert 0\right\rangle \langle 0|=\colon e^{-a^{\dagger
}a}\colon $ and noticing that $a^{\dagger }$ is commutable with $a$ within
normal ordering symbol $\colon \ \colon ,$ we can perform the summation in (%
\ref{9}) 
\begin{eqnarray}
\rho \left( t\right)  &=&\int \frac{d^{2}\alpha }{\pi }e^{-|\alpha
|^{2}}P(\alpha ,0)\sum_{m,n=0}^{\infty }\frac{\left\vert \alpha \right\vert
^{2n}}{m!n!}\frac{\left( \kappa t\right) ^{m+n}}{\left( \kappa t+1\right)
^{m+n+1}}\colon a^{\dagger m}a^{m}e^{\left( \alpha a^{\dag }+\alpha ^{\ast
}a\right) \frac{1}{1+\kappa t}-a^{\dag }a}\colon   \notag \\
&=&\int \frac{d^{2}\alpha }{\pi }e^{-|\alpha |^{2}}P(\alpha
,0)\sum_{m=0}^{\infty }\frac{1}{m!}\frac{\left( \kappa t\right) ^{m}}{\left(
\kappa t+1\right) ^{m+1}}e^{\frac{\kappa t}{\kappa t+1}\left\vert \alpha
\right\vert ^{2}}\colon a^{\dagger m}a^{m}e^{\left( \alpha a^{\dag }+\alpha
^{\ast }a\right) \frac{1}{1+\kappa t}-a^{\dag }a}\colon   \label{10} \\
&=&\frac{1}{\kappa t+1}\int \frac{d^{2}\alpha }{\pi }e^{-\frac{|\alpha |^{2}%
}{\kappa t+1}}P(\alpha ,0)\colon e^{\frac{\kappa t}{\kappa t+1}a^{\dagger }a+%
\frac{1}{1+\kappa t}\left( \alpha a^{\dag }+\alpha ^{\ast }a\right) -a^{\dag
}a}\colon .  \notag
\end{eqnarray}%
Once the $P$-representation $P(\alpha ,0)$ is known, we can derive $\rho
\left( t\right) $ by directly performing the integration in (\ref{10}) by
virtue of the IWOP method. For an initial coherent state $\rho
_{0}=|z\rangle \langle z|,$ its $P$-representation is Delta function 
\begin{equation*}
P(\alpha ,0)_{|z\rangle \langle z|}=\pi \delta \left( \alpha ^{\ast
}-z^{\ast }\right) \delta \left( \alpha -z\right) ,
\end{equation*}%
then from Eq. (\ref{10}) we immediately have the final state in the
diffusion channel%
\begin{eqnarray}
\rho \left( t\right) _{|z\rangle \langle z|} &=&\frac{1}{\kappa t+1}%
:e^{-|z|^{2}\frac{1}{\kappa t+1}+\frac{\kappa t}{\kappa t+1}a^{\dagger
}a+\left( za^{\dag }+z^{\ast }a\right) \frac{1}{1+\kappa t}-a^{\dag }a}:
\label{11} \\
&=&\frac{1}{\kappa t+1}e^{-|z|^{2}\frac{1}{\kappa t+1}}e^{za^{\dag }\frac{1}{%
1+\kappa t}}e^{a^{\dagger }a\ln \frac{\kappa t}{\kappa t+1}}e^{z^{\ast }a%
\frac{1}{1+\kappa t}},  \notag
\end{eqnarray}%
which is a mixed state involving a chaotic light characteristic of $%
e^{a^{\dagger }a\ln \frac{\kappa t}{\kappa t+1}}$, thus we conclude that
through a diffusion process a pure coherent state evolves into a mixed
state, and we can confirm that Eq. (\ref{11}) is qualified being a density
operator as we can prove 
\begin{equation}
\text{Tr}\left[ \rho \left( t\right) _{|z\rangle \langle z|}\right] =\int 
\frac{d^{2}\alpha }{\pi }\left\langle \alpha \right\vert \rho \left(
t\right) _{|z\rangle \langle z|}|\alpha \rangle =\frac{1}{\kappa t+1}%
e^{-|z|^{2}\frac{1}{\kappa t+1}}\int \frac{d^{2}\alpha }{\pi }e^{-\frac{%
|\alpha |^{2}}{\kappa t+1}+\left( z\alpha ^{\ast }+z^{\ast }\alpha \right) 
\frac{1}{1+\kappa t}}=1.  \label{12}
\end{equation}%
Then we can further prove the classical correspondence of $\rho \left(
t\right) _{|z\rangle \langle z|}$ really obey the classical diffusion
equation, for this aim, we use the operator identity which can turn a
density operator into its antinormal ordering form \cite{Mehta_1967_prl} 
\begin{equation}
\rho =\int \frac{d^{2}\beta }{\pi }\vdots \langle -\beta |\rho |\beta
\rangle \exp [|\beta |^{2}+\beta ^{\ast }a-\beta a^{\dag }+a^{\dag }a]\vdots
,  \label{13}
\end{equation}%
where $\vdots $ $\vdots $ denotes anti-normal ordering \cite{Fan_1991_pla,
Fan_1992_cjqe}, substituting (\ref{11}) into the right-hand side of (\ref{13}%
) we have 
\begin{eqnarray}
\rho _{|z\rangle \langle z|}(t) &=&\frac{1}{1+\kappa t}\int \frac{d^{2}\beta 
}{\pi }\vdots \exp [-\frac{\kappa t}{1+\kappa t}|\beta |^{2}+\beta ^{\ast
}(a-\frac{z}{1+\kappa t})+\beta (\frac{z^{\ast }}{1+\kappa t}-a^{\dag })-%
\frac{|z|^{2}}{1+\kappa t}+a^{\dag }a]\vdots   \notag \\
&=&\frac{1}{\kappa t}\vdots \exp [-\frac{1}{\kappa t}(z-a)(z^{\ast }-a^{\dag
})]\vdots ,  \label{14}
\end{eqnarray}%
this is the anti-normally ordered form of $\rho \left( t\right) _{|z\rangle
\langle z|}.$ Its classical representation in the coherent state basis is 
\begin{equation}
\rho \left( t\right) _{|z\rangle \langle z|}=\int \frac{d^{2}\alpha }{\pi }%
P\left( \alpha ,t\right) |\alpha \rangle \langle \alpha |,  \label{15}
\end{equation}%
with%
\begin{equation}
P\left( \alpha ,t\right) =\frac{1}{\kappa t}\exp [-\frac{1}{\kappa t}%
(z-\alpha )(z^{\ast }-\alpha ^{\ast })].  \label{16}
\end{equation}%
We can check that it really obeys the classical diffusion equation%
\begin{equation}
\frac{\partial P\left( \alpha ,t\right) }{dt}=\kappa \frac{\partial ^{2}}{%
\partial \alpha \partial \alpha ^{\ast }}P\left( \alpha ,t\right) .
\label{17}
\end{equation}%
For the origin of this equation we refer to the Appendix of this paper.

\section{Integration-form solution to the dissipative master equation (form
2)}

To go a step further, using the inverse relation of the $P$-representation
in Eq. (\ref{7})%
\begin{equation}
P(\alpha ,0)=e^{|\alpha |^{2}}\int \frac{d^{2}\beta }{\pi }\left\langle
-\beta \right\vert \rho _{0}\left\vert \beta \right\rangle e^{|\beta
|^{2}+\beta ^{\ast }\alpha -\beta \alpha ^{\ast }},  \label{18}
\end{equation}%
where $\left\vert \beta \right\rangle $ is also a coherent state \ref%
{Klauder_1985_cs}\ and substituting Eq. (\ref{14}) into Eq. (\ref{10}) yields%
\begin{eqnarray}
\rho \left( t\right)  &=&\frac{1}{\kappa t+1}\int \frac{d^{2}\beta }{\pi }%
\left\langle -\beta \right\vert \rho _{0}\left\vert \beta \right\rangle
e^{|\beta |^{2}}\int \frac{d^{2}\alpha }{\pi }e^{\frac{\kappa t|\alpha |^{2}%
}{\kappa t+1}}e^{\beta ^{\ast }\alpha -\beta \alpha ^{\ast }}\colon e^{\frac{%
1}{1+\kappa t}\left( \alpha a^{\dag }+\alpha ^{\ast }a\right) -\frac{1}{%
\kappa t+1}a^{\dagger }a}\colon   \notag \\
&=&-\frac{1}{\kappa t}\int \frac{d^{2}\beta }{\pi }\left\langle -\beta
\right\vert \rho _{0}\left\vert \beta \right\rangle e^{|\beta |^{2}}  \notag
\\
&&\times \colon \exp \left[ \left( \frac{\kappa t+1}{-\kappa t}\right)
\left( \frac{a^{\dag }}{1+\kappa t}+\beta ^{\ast }\right) \left( \frac{a}{%
1+\kappa t}-\beta \right) -\frac{1}{\kappa t+1}a^{\dagger }a\right] \colon 
\label{19} \\
&=&\frac{-1}{\kappa t}\int \frac{d^{2}\beta }{\pi }\left\langle -\beta
\right\vert \rho _{0}\left\vert \beta \right\rangle e^{|\beta |^{2}}\colon
\exp \left\{ \frac{1}{\kappa t}\left[ |\beta |^{2}\left( \kappa t+1\right)
+\beta a^{\dag }-\beta ^{\ast }a-a^{\dagger }a\right] \right\} \colon , 
\notag
\end{eqnarray}%
this is the integration-form solution to the diffusion master equation,
connecting input state $\rho _{0}$ with its output state $\rho \left(
t\right) $.

\section{Applications}

Eq. (\ref{15}) brings much convenience for obtaining the time evolution law
in the diffusion process. As an example, we consider the case of an initial
number state $\left \vert l\right \rangle \left \langle l\right \vert $
undergoing through the diffusion channel, here $\left \vert l\right \rangle
=a^{\dagger n}\left \vert 0\right \rangle /\sqrt{n!}.$ Due to 
\begin{equation}
\left \langle l\right \vert \left. \beta \right \rangle =e^{-|\beta |^{2}/2}%
\frac{\beta ^{l}}{\sqrt{l!}}  \label{20}
\end{equation}%
and%
\begin{equation}
\left \langle -\beta \right. |l\rangle \left \langle l\right. \left \vert
\beta \right \rangle =\frac{\left( -1\right) ^{l}|\beta |^{2l}}{l!}%
e^{-|\beta |^{2}}.  \label{21}
\end{equation}%
Substituting (\ref{16}) into (\ref{14}) and using the formula 
\begin{equation}
\int \frac{d^{2}\beta }{\pi }\beta ^{n}\beta ^{\ast m}\exp \left( \zeta
\left \vert \beta \right \vert ^{2}+\xi \beta +\eta \beta ^{\ast }\right)
=e^{\frac{-\xi \eta }{\zeta }}\sum_{k=0}^{\min \left[ n,m\right] }\frac{%
n!m!\xi ^{m-k}\eta ^{n-k}}{k!\left( n-k\right) !\left( m-k\right) !\left(
-\zeta \right) ^{m+n-k+1}}.  \label{22}
\end{equation}%
Using the definition of the two-variable Hermite polynomials%
\begin{equation}
H_{m,n}\left( x,y\right) =\sum_{l=0}^{\min (m,n)}\frac{m!n!(-1)^{l}}{%
l!(m-l)!(n-l)!}x^{m-l}y^{n-l},  \label{23}
\end{equation}%
as well as the definition of Laguerre polynomials%
\begin{equation}
L_{l}\left( x\right) =\sum \binom{l}{l-k}\frac{\left( -x\right) ^{k}}{k!}
\label{24}
\end{equation}%
and%
\begin{equation}
L_{l}\left( xy\right) =\frac{\left( -1\right) ^{l}}{l!}H_{l,l}(x,y),
\label{25}
\end{equation}%
we can derive%
\begin{eqnarray}
\rho \left( t\right) &=&\frac{-1}{\kappa t}\int \frac{d^{2}\beta }{\pi }%
\left \langle -\beta \right \vert \left. \beta \right \rangle \left \langle
l\right \vert \left. \beta \right \rangle e^{|\beta |^{2}}\colon \exp \left
\{ \frac{1}{\kappa t}\left[ |\beta |^{2}\left( \kappa t+1\right) +\beta
a^{\dag }-\beta ^{\ast }a-a^{\dagger }a\right] \right \} \colon  \notag \\
&=&\frac{\left( -1\right) ^{l+1}}{l!\kappa t}\int \frac{d^{2}\beta }{\pi }%
|\beta |^{2l}\colon \exp \left \{ \frac{1}{\kappa t}\left[ |\beta
|^{2}\left( \kappa t+1\right) +\beta a^{\dag }-\beta ^{\ast }a-a^{\dagger }a%
\right] \right \} \colon  \label{26} \\
&=&\frac{\left( \kappa t\right) ^{l}}{\left( \kappa t+1\right) ^{l+1}}\colon
L_{l}\left( \frac{-a^{\dagger }a}{\kappa t\left( \kappa t+1\right) }\right)
e^{\frac{-1}{\kappa t+1}a^{\dagger }a}\colon ,  \notag
\end{eqnarray}%
which is named Laguerre-polynomial-weighted chaotic state, since $e^{\frac{-1%
}{\kappa t+1}a^{\dagger }a}$ represents a chaotic photon field, here the
symbol "$\colon \colon $"\ denotes normal ordering symbol. Thus we see $%
\left \vert l\right \rangle \left \langle l\right \vert $ evolves into the
mixed state (\ref{29}), so this diffusion process manifestly embodies
quantum decoherence. We can further calculate the average photon number $Tr%
\left[ \rho \left( t\right) a^{\dagger }a\right] =l+\kappa t$, which tells
that in the diffusion process, the photon number $l\rightarrow l+\kappa t.$

As the second example, If the initial state is a squeezed state with a
squeezing parameter $\lambda $, whose density operator is given by%
\begin{equation}
\rho _{0}=\text{sech}\lambda \text{ }e^{\frac{1}{2}a^{\dagger 2}\tanh
\lambda }|0\rangle \langle 0|e^{\frac{1}{2}a^{2}\tanh \lambda }.  \label{27}
\end{equation}%
Substituting (\ref{22}) into (\ref{14}) and using%
\begin{equation}
\left \langle -\beta \right \vert \rho _{0}\left \vert \beta \right \rangle =%
\text{sech}\lambda e^{\frac{1}{2}\left( \beta ^{\ast 2}+\beta ^{2}\right)
\tanh \lambda -|\beta |^{2}},  \label{28}
\end{equation}%
as well as the integration formula%
\begin{equation}
\int \frac{d^{2}z}{\pi }\exp \left( \zeta \left \vert z\right \vert ^{2}+\xi
z+\eta z^{\ast }+fz^{2}+gz^{\ast 2}\right) =\frac{1}{\sqrt{\zeta ^{2}-4fg}}%
\exp \left[ \frac{-\zeta \xi \eta +f\eta ^{2}+g\xi ^{2}}{\zeta ^{2}-4fg}%
\right] ,  \label{29}
\end{equation}%
we can finally obtain a mixed state%
\begin{eqnarray}
\rho \left( t\right) &=&-\frac{\text{sech}\lambda }{\kappa t}\int \frac{%
d^{2}\beta }{\pi }\colon \exp \left \{ \frac{\kappa t+1}{\kappa t}|\beta
|^{2}+\frac{\beta a^{\dag }-\beta ^{\ast }a}{\kappa t}+\frac{\tanh \lambda }{%
2}\left( \beta ^{\ast 2}+\beta ^{2}\right) -\frac{a^{\dagger }a}{\kappa t}%
\right \} \colon  \notag \\
&=&-\text{sech}\lambda \sqrt{\frac{1}{G}}\colon \exp \left[ \frac{\frac{%
\kappa t+1}{\kappa t}a^{\dagger }a+\frac{\tanh \lambda }{2}\left( a^{\dagger
2}+a^{2}\right) }{G}-\frac{a^{\dagger }a}{\kappa t}\right] \colon  \label{30}
\\
&=&-\text{sech}\lambda \sqrt{\frac{1}{G}}e^{\frac{\tanh \lambda }{2G}%
a^{\dagger 2}}\colon \exp \left[ \frac{\left( \kappa t+1\right) a^{\dagger }a%
}{G\kappa t}-\frac{a^{\dagger }a}{\kappa t}\right] \colon e^{\frac{\tanh
\lambda }{2G}a^{2}},  \notag
\end{eqnarray}%
which is a mixed state, where 
\begin{equation*}
G\equiv \left( \kappa t+1\right) ^{2}-\left( \kappa t\right) ^{2}\tanh
^{2}\lambda .
\end{equation*}%
Actually, Eq.(\ref{24}) can be considered as a squeezed thermal state.

In summary, for master equation describing the dissipative channel, we
developed the form solution of Kraus into a new integration form by using
the IWOP technique. It will be more concise to derive various density
operator $\rho \left( t\right) $ and can help us to understand deeply the
quantum decoherence and reduce calculation vastly.

\section{Appendix}

\bigskip We now demonstrate that the classical correspondence to the
diffusion master equation%
\begin{equation}
\frac{d\rho }{dt}=-\kappa (a^{\dag }a\rho -a^{\dag }\rho a-a\rho a^{\dag
}+\rho aa^{\dag })  \label{A1}
\end{equation}%
is Eq. (\ref{17}). In fact, using%
\begin{equation}
\rho \left( t\right) =\int \frac{d^{2}\alpha }{\pi }P(\alpha ,t)|\alpha
\rangle \langle \alpha |,  \label{A2}
\end{equation}%
Eq. (\ref{A1}) becomes to

\begin{equation}
\frac{d\rho }{dt}=-\kappa \int \frac{d^{2}\alpha }{\pi }P(\alpha ,t)(a^{\dag
}a|\alpha \rangle \langle \alpha |-a^{\dag }|\alpha \rangle \langle \alpha
|a-a|\alpha \rangle \langle \alpha |a^{\dag }+|\alpha \rangle \langle \alpha
|aa^{\dag }).  \label{A3}
\end{equation}%
Considering

\begin{eqnarray}
a^{\dag }|\alpha \rangle \langle \alpha | &=&\colon a^{\dag }\colon
e^{-|\alpha |^{2}+\alpha a^{\dag }+\alpha ^{\ast }a-a^{\dag }a}\colon
=(\alpha ^{\ast }+\frac{\partial }{\partial \alpha })|\alpha \rangle \langle
\alpha |,  \notag \\
|\alpha \rangle \langle \alpha |a &=&(\alpha +\frac{\partial }{\partial
\alpha ^{\ast }})|\alpha \rangle \langle \alpha |,  \label{A4}
\end{eqnarray}%
we can see

\begin{eqnarray}
&&a^{\dag }a|\alpha \rangle \langle \alpha |-a^{\dag }|\alpha \rangle
\langle \alpha |a-a|\alpha \rangle \langle \alpha |a^{\dag }+|\alpha \rangle
\langle \alpha |aa^{\dag }  \notag \\
&=&\alpha a^{\dag }|\alpha \rangle \langle \alpha |-(\alpha ^{\ast }+\frac{%
\partial }{\partial \alpha })|\alpha \rangle \langle \alpha |a-|\alpha
|^{2}|\alpha \rangle \langle \alpha |+(\alpha +\frac{\partial }{\partial
\alpha ^{\ast }})|\alpha \rangle \langle \alpha |a^{\dag }  \notag \\
&=&\alpha (\alpha ^{\ast }+\frac{\partial }{\partial \alpha })|\alpha
\rangle \langle \alpha |-(\alpha ^{\ast }+\frac{\partial }{\partial \alpha }%
)(\alpha +\frac{\partial }{\partial \alpha ^{\ast }})|\alpha \rangle \langle
\alpha |-|\alpha |^{2}|\alpha \rangle \langle \alpha |+(\alpha +\frac{%
\partial }{\partial \alpha ^{\ast }})(\alpha ^{\ast }|\alpha \rangle \langle
\alpha |)  \label{A5} \\
&=&-\frac{\partial ^{2}}{\partial \alpha \partial \alpha ^{\ast }}|\alpha
\rangle \langle \alpha |.  \notag
\end{eqnarray}%
Inserting (\ref{A5}) into (\ref{A3}), we obtain%
\begin{equation}
\frac{d\rho }{dt}=\kappa \int \frac{d^{2}\alpha }{\pi }\frac{\partial
^{2}P(\alpha ,t)}{\partial \alpha \partial \alpha ^{\ast }}|\alpha \rangle
\langle \alpha |.  \label{A6}
\end{equation}%
On the other hand, we have%
\begin{equation}
\frac{d\rho }{dt}=\int \frac{d^{2}\alpha }{\pi }\frac{\partial P(\alpha ,t)}{%
\partial t}|\alpha \rangle \langle \alpha |.  \label{A7}
\end{equation}%
Comparing (\ref{A6}) with (\ref{A7}) we derive Eq. (\ref{17}), the classical
diffusion equation.


\begin{thebibliography}{99}
\bibitem{Preskill_1998_lnp} J. Preskill, \emph{Quantum Information and
Computation}, Lecture Notes for Physics, Vol. 229 (CIT) (1998).

\bibitem{Gardiner_2000_qn} C. Gardiner and P. Zoller, \emph{Quantum Noise}
(Springer, Berlin) (2000).

\bibitem{zhang_2013_aps} H. L. Zhang, F. Jia, X. X. Xu, Q. Guo, X. Y. Tao
and L. Y. Hu, \textit{Acta Phys. Sin.} \textbf{6}2 ( 2013) 014208.

\bibitem{Carmichael_1999_mefp} H. J. Carmichael, \emph{Statistical Methods
in Quantum Optics 1: Master Equations and Fokker-Planck Equations},
(Springer-Verlag, Berlin) (1999).

\bibitem{Carmichael_2008_ncf} H. J. Carmichael, \emph{Statistical Methods in
Quantum Optics 2: Non-Classical Fields}, (Springer-Verlag, Berlin) (2008).

\bibitem{Kraus_1983_lnp} K. Kraus, \emph{States, Effects, and Operations:
Fundamental Notions of Quantum Theory}, Lecture Notes in Physics, Vol. 190
(Springer-Verlag, Berlin) (1983)

\bibitem{Fan_2008_mplb} H. Y. Fan and L. Y. Hu, \emph{Mod. Phys. Lett. B} 
\textbf{22}, 2435 (2008).

\bibitem{Klauder_1985_cs} See e.g., J. R. Klauder and B. S. Skargerstam, 
\emph{Coherent States}, (World Scientific, Singapore) (1985).

\bibitem{Fan_2003_jopb} H. Y. Fan \emph{J. Opt. B: Quantum Semiclass. Opt.} 
\textbf{5,} R147 (2003).

\bibitem{Sudanshan_1963_prl} E. C. G. Sudanshan, \emph{Phys. Rev. Lett. }%
\textbf{10,} 277 (1963)

\bibitem{Fan_1991_pla} H. Y. Fan, \emph{Phys. Lett. A} \textbf{131}(3) 145
(1988); \textbf{161}(1), 1 (1991).

\bibitem{Fan_1992_cjqe} H. Y. Fan, H. G. Weng, \emph{Chinese Journal of
Quantum Electronics} \textbf{9}(3), 213 (1992).

\bibitem{Mehta_1967_prl} C. L. Mehta \textit{Phys. Rev. Lett}. \textbf{18,}
752 (1967).
\end{thebibliography}
\end{document}